\documentclass[%
12pt,
  reprint,
 superscriptaddress,
 amsmath,amssymb,
 aps,UTF8,
]{revtex4-1}

\usepackage{footmisc}
\usepackage{graphicx}
\usepackage{dcolumn}
\usepackage{bm}
\usepackage{xcolor}
\usepackage{url}
\usepackage{float}
\usepackage{tabularx}
\usepackage{lipsum}
\usepackage{array}

\usepackage{booktabs}

\usepackage{subfigure}
\usepackage{multirow}

\begin{document}


\title{Exploring quantum entanglement in chiral symmetry partial restoration with 1+1 string model}
\author{Wei Kou}
\email{kouwei@impcas.ac.cn}
\affiliation{Institute of Modern Physics, Chinese Academy of Sciences, Lanzhou 730000, China}
\affiliation{University of Chinese Academy of Sciences, Beijing 100049, China}
\author{Xurong Chen}
\email{xchen@impcas.ac.cn (Corresponding author)}
\affiliation{Institute of Modern Physics, Chinese Academy of Sciences, Lanzhou 730000, China}
\affiliation{University of Chinese Academy of Sciences, Beijing 100049, China}
	\affiliation{Southern Center for Nuclear Science Theory (SCNT), Institute of Modern Physics, Chinese Academy of Sciences, Huizhou 516000, Guangdong Province, China}


\begin{abstract}
Within the confining color flux tube picture, we assess the color electric field generated by quark-antiquark pairs within the framework of the Schwinger model and estimate its impact on the chiral condensate. We observe differences in the distribution of the color flux tube generated by quark-antiquark pairs at different separation distances, leading to discrepancies in the partial restoration of chiral symmetry. Furthermore, we suggest incorporating the color magnetic field in the calculation of chiral condensate, leading to quantum entanglement effects, and proceed to compute the entanglement entropy. We observe that the entanglement entropy increases with the distance of the color source (the separation distance between quark and anti-quark), and the magnitude of the color electric field and chiral condensate restoration after spatial integration also increases with the distance of the color source. Then, we try to provide a qualitative explanation for the existence of these phenomena.

\end{abstract}

\maketitle


\section{Introduction}
\label{sec:intro}
The Schwinger model \cite{Schwinger:1962tn,Schwinger:1962tp} with massless fermions is exactly solvable in (1+1) dimensional spacetime. From this classical solvable model, one can gain insight into some properties of Quantum Chromodynamics (QCD), such as chiral symmetry breaking and confinement (nonzero fermion mass). These unresolved issues motivate physicists to deeply explore QCD, and much related works have been done by using the equivalence of the massless fermion case of the Schwinger model to the theory of free massive boson fields \cite{Brown:1963po,Zumino:1964nja,Lowenstein:1971fc,Casher:1974vf,Coleman:1975pw,Kharzeev:2014xta,Tomiya:2022chr,Florio:2023dke}.

As one of the unsolved mysteries in QCD, the quark confinement problem has consistently troubled many researchers. In recent years, researchers have made significant progress in confinement studies by using the description of the Schwinger model in holographic QCD to establish a connection between the quantum entanglement of quark-antiquark pairs accelerating away from each other in the vacuum and the Einstein-Rosen bridge, and incorporating confinement potential into the analysis \cite{Xiao:2008nr,Semenoff:2011ng,Chernicoff:2013iga,Jensen:2013ora,Lewkowycz:2013laa,Sonner:2013mba,Hubeny:2014zna,Jensen:2014bpa,Ghodrati:2015rta,Grieninger:2023ehb,Grieninger:2023pyb}. These works inspire the question: Is there a duality between wormholes in the bulk and quantum entanglement? 

In general, the interpretation of the confinement effect in the language of string theory can be simplified as the breaking of strings due to the separation of quark-antiquark pairs, leading to the excitation of new quark-antiquark pairs from the vacuum. The newly generated quark-antiquark pairs can affect the distribution of color-electric fields in the vacuum, thus influencing the vacuum dynamics of QCD. Discussions of such phenomena have been reflected in recent simulation calculations \cite{Florio:2023dke,Ikeda:2023zil}, still based on the Schwinger model.

Another feature of QCD is the chiral symmetry breaking. As massless quarks move in the QCD vacuum, they acquire dynamical mass, resulting in the emergence of a non-zero quark condensate $\langle q\bar{q} \rangle_0$. In other words, chiral symmetry breaking does not occur in the presence of chiral symmetry. This property is referred to as dynamical chiral symmetry spontaneous breaking (DCSB) \cite{Nambu:1961tp,Nambu:1961fr}. It is meaningful to calculate the vacuum chiral condensate using an appropriate model. Ref. \cite{Hamer:1982mx} provides a method for computing the condensate considering the massive Schwinger model. Assuming the change in vacuum condensate behavior under the influence of background fields is interesting, which has prompted the development of a theoretical framework to describe non-Abelian background fields in QCD. The description of (3+1)-dimensional non-Abelian gauge fields is complex. In order to simplify the framework, a quasi-Abelian picture based on the dual Meissner effect has been proposed, providing support for many lattice QCD studies \cite{Singh:1993jj,Schilling:1998gz,Chernodub:2000rg,Chernodub:2005gz,Suzuki:2009xy,Cea:2012qw}. This is also referred to as the ``maximum Abelian projection." Thees arguments are summarized and reviewed in Ref. \cite{Kharzeev:2014xta}, where the relationship between confinement and vacuum magnetic monopole condense is discussed. In a simple string model, confined quarks are connected by Abelian color-electric flux tubes. These flux tubes have thin transverse size, and are dual to Abrikosov-Nielsen-Olesen (ANO) vortices \cite{Abrikosov:1956sx,Nielsen:1973cs,Cea:2012qw} in type-II superconductors. This implies that charged massless fermions can be localized to the core of ANO vortices containing a magnetic field, and their dynamical behavior can be described by (1+1) dimensional effective theory \cite{Witten:1984eb}. 

The difference between vacuum chiral condensate and chiral condensate in a background field is intriguing, and how background fields affect the vacuum properties of QCD is a worthwhile question to explore. We assume that the chiral condensate,  generated by chiral symmetry spontaneous breaking in vacuum, can be analogized as a Hilbert space quantum pure state, and its value can be computed using specific models. When a background field is introduced, whether it is a non-Abelian gauge field or a quasi-Abelian gauge field (currently only dealing with the Abelian case), it breaks the ``stable" vacuum condensate and causes a decrease in the condensate value, known as the chiral symmetry partial restoration \cite{Kharzeev:2014xta,Kaneko:2013jla,Iritani:2013apa,Iritani:2013rla,Iritani:2014fga,Iritani:2015zwa,Iritani:2016fvi}.  The changes induced by the background field on chiral condensates may be reflected in quantum entanglement, which should be sourced by the background field (quark-antiquark pairs). In this work, we employ the (1+1)-dimensional Schwinger model to construct the background field and calculate the variation of the vacuum chiral condensate under the background field. Additionally, we introduce two subspaces of the total Hilbert space after the direct product decomposition: (i) the chiral condensate space without the background field and (ii) the chiral condensate space affected by the background field. Then we calculate the entanglement entropy. We change the distance between the quark-antiquark pairs, affecting the distribution of the background field, and find that the entanglement entropy increases with the increased distance between the quark-antiquark pairs, similar to recent quantum simulation results. We argue that in the framework of the (1+1)-dimensional Schwinger model, the entanglement entropy between the quark-antiquark pairs will increase with their distance, and as the source of the background field, the entanglement entropy between them will be reflected in the vacuum chiral condensate they affect.

The sections organization of this work are as follows. In section \ref{sec:theory}, we briefly review some concepts related to the partial restoration of chiral symmetry, followed by the introduction of entanglement entropy. In section \ref{sec:dissc}, we present the computational results and discuss possible correspondences between several physical quantities. In the final section, we provide the conclusion and outlook for the entire paper.

\section{Formalism}
\label{sec:theory}
\subsection{1+1 Schwinger model and chiral condensate}
\label{subsec:1+1}
This section begins with the construction of the chiral condensate using the bosonization of the Schwinger model. The Lagrangian of the Abelian gauge theory in (1+1) dimensions for the case of massless fermions is given by
\begin{equation}
	\mathcal{L}=-\frac14F_{\mu\nu}F^{\mu\nu}+\bar{\psi}(i\gamma^\mu\partial_\mu-g\gamma^\mu A_\mu)\psi,
	\label{eq:Lagrangian}
\end{equation}
where $g$ is the coupling constant with the dimension of mass, $F_{\mu\nu}$ is gauge strength, $\psi$ denotes the fermion field. As mentioned earlier, this theory is exactly solvable and processes properties such as confinement, chiral symmetry breaking, axial anomaly, and periodic $\theta$-vacuum \cite{Schwinger:1962tp,Lowenstein:1971fc,Coleman:1975pw}. Additionally, it is used to calculate the chiral condensate depended on the electric field \cite{Hamer:1982mx}. In other words, the presence of a background electric field suppresses the magnitude of the chiral condensate, with its periodicity depending on the $\theta$-vacuum of the theoretical model.

In one-dimensional space, fermionic degrees of freedom can be precisely expressed in terms of bosonic degrees of freedom. This correspondence is typically manifested as duality between fermionic bilinears and the real scalar fields. This scheme is known as bosonization. In the case of an Abelian gauge, the correspondence is as follows \cite{Coleman:1974bu,Mandelstam:1975hb}
\begin{equation}
	\begin{aligned}
		&\bar{\psi}i\gamma^\mu\partial_\mu\psi\to\frac12\partial_\mu\phi\partial^\mu\phi,\\
		&\bar{\psi}\gamma^\mu\psi\to-\frac1{\sqrt{\pi}}\epsilon^{\mu\nu}\partial_\nu\phi,\\
		&\bar{\psi}\psi\to-c\frac g{\sqrt{\pi}}\cos(2\sqrt{\pi}\phi),
	\end{aligned}
	\label{eq:bose}
\end{equation}
where $c=e^{\gamma_{E}}/2\pi$ with the Euler constant $\gamma_E\simeq0.5772$. Note that the scalar field has mass dimension $[\phi]=0$. Introducing the effective mass of scalar field, the Lagrangian (\ref{eq:Lagrangian}) is equivalent to
\begin{equation}
	\mathcal{L}=\frac12\partial_\mu\phi\partial^\mu\phi-\frac12\frac{g^2}\pi\phi^2.
	\label{eq:Lagrangian-phi}
\end{equation}
From Eq. (\ref{eq:bose}), one can express the chiral condensate in terms of the scalar filed is $\bar{\psi}\psi=-\frac{ge^{\gamma_{E}}}{2\pi^{3/2}}\cos(2\sqrt{\pi}\phi)$. Thus the chiral condensate is evaluated through the Feynman-Hellman theorem by differentiating the energy of the ground state in the presence of an electric filed $E^{1+1}$ with respect to the fermion mass $m$, in the chiral limit $m\to 0$ \cite{Hamer:1982mx}:
\begin{equation}
	\langle\bar{\psi}\psi\rangle_{E_{1+1}}=-\frac{ge^{\gamma_E}}{2\pi^{3/2}}\cos\theta,
	\label{eq:chiral condensate}
\end{equation}
with $\theta=2\pi E_{1+1}/g$. Here, $E_{1+1}$ is the electric field along the string flux tube. Formally, one can see that the value of the condensate becomes a constant once the background electric field is specified. If the electric field $E_{1+1}=0$, the condensate must be constant $\langle\bar{\psi}\psi\rangle_0=-\frac{ge^{\gamma_E}}{2\pi^{3/2}}$ with the coupling $g$.

\subsection{3+1 dimensional physical electric field and string fluctuation}
\label{subsec:3+1}
\begin{figure}[htbp]
	\centering
	\includegraphics[width=0.48\textwidth]{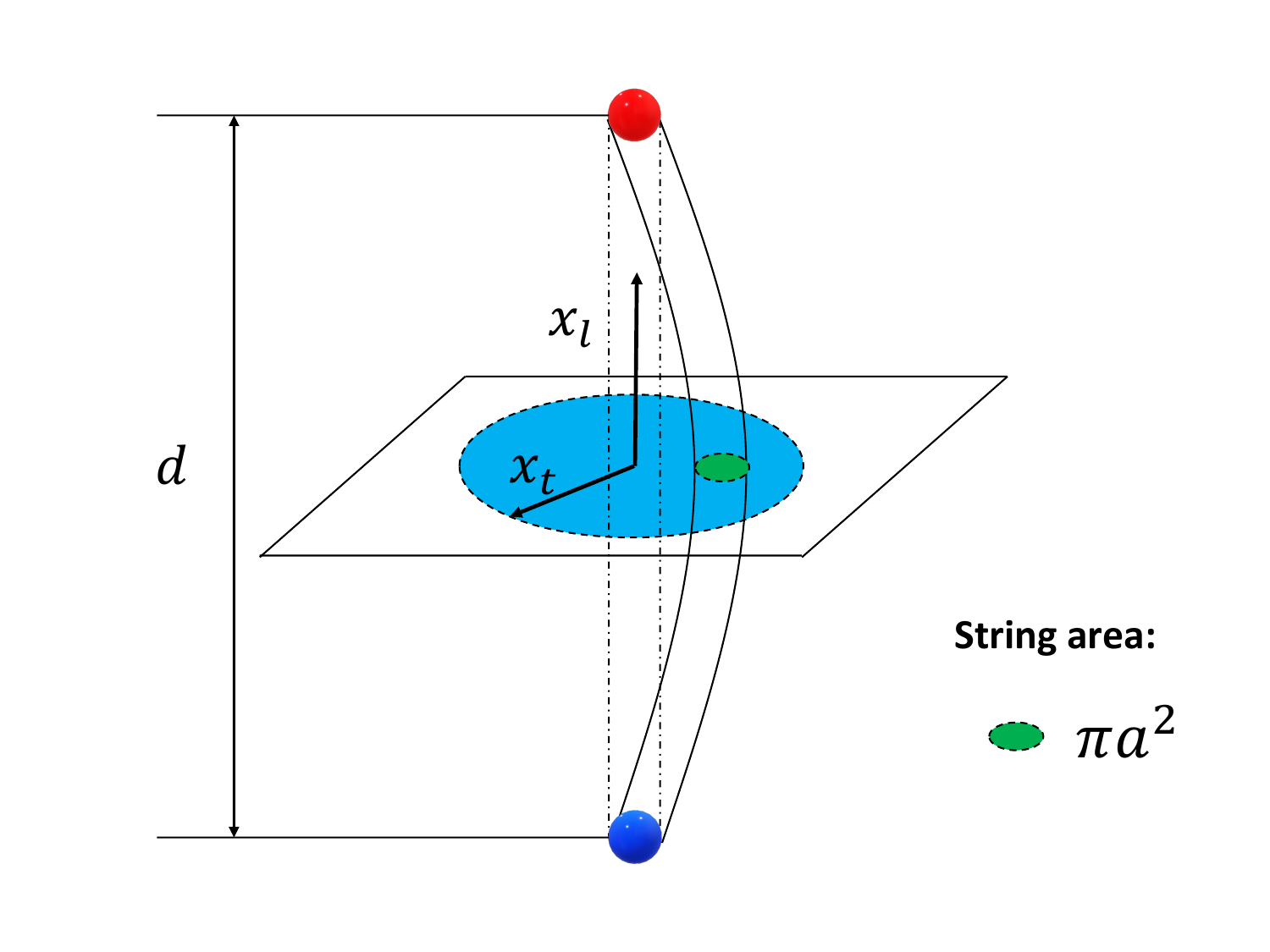}
	\caption{The schematic diagram of the electric field string flux tube. Positive and negative static fermions (separated by a longitudinal distance $d$) are connected by a thin color flux tube with width $a$, the string fluctuating in the transverse space ($x_t$ direction). The string area is simplified to $\pi a^2$.}
	\label{fig:string}
\end{figure}

The actual physical presence of a (3+1)-dimensional electric field distribution, with the ``thin" (1+1)-dimensional string corresponding to the electric field along the flux tube direction ($x_l$ direction), see Figure \ref{fig:string}. The electric field present in reality should have complete physical significance, denoted as $E_{3+1}(x_t)$, where $x_t$ is the transverse coordinate in the plane perpendicular to the flux tube. In this scenario, it has been demonstrated that the color fields of a static quark-antiquark pair are almost entirely described by the longitudinal chromoelectric field, which in turn is approximately constant along the flux tube \cite{Cea:1995zt,Cardaci:2010tb}. Therefore, one can assume that $E_{1+1}$ and $E_{3+1}(x_t)$ have the same string tension in the longitudinal flux direction, i.e., the energy per unit length of the string should be equal \cite{Cea:2012qw,Kharzeev:2014xta}:
\begin{equation}
	\sigma_{E}\simeq\frac12\int d^2x_tE_{3+1}^2(x_t)=\frac12(E_{1+1})^2.
	\label{eq:3+1=1+1}
\end{equation}
Clearly, the above equation allows one to introduce a quantity $P(x_t)$ to represent the probability of finding fluctuations of the string at the transverse plane position $x_t$, and the probability distribution is constrained as:
\begin{equation}
	(E_{3+1}(x_t))^2=(E_{1+1})^2 P(x_t),
	\label{eq:P(xt)}
\end{equation}
since $\int d^2x_t P(x_t)=1$ condition is satisfied.

Figure \ref{fig:string} depicts the transverse plane position fluctuations of the (1+1)-dimensional string. Although it is not possible to accurately describe this fluctuation process, the information of the fluctuations is encoded into the introduced probability $P(x_t)$. This is reasonable under the assumption that Eq. (\ref{eq:3+1=1+1}) holds. 

\subsection{Partial restoration of chiral symmetry}
Let us now consider how the vacuum chiral condensate will change under the influence of the background field. Firstly, a natural way to describe the variation of chiral symmetry in the background field is the ratio of the condensate with and without the background field. It is expected that the ratio should be 1.0 when there is no background electric field, which is obvious. When the background field is present, as described in the previous subsection, the non-zero $E_{1+1}$ should be included in the calculation of the condensate (\ref{eq:chiral condensate}). Here we need to consider the property of the transverse plane fluctuations of the string described in Figure \ref{fig:string}, and take into account the size $a$ of the string and its corresponding effective area $\pi a^2$. Therefore, the probability of finding the string (flux) at a given transverse position is $\pi a^2 P(x_t)$, indicating that if we want to find the electric field flux tube at a certain $x_t$, the condensate will incorporate $x_t$-dependent information \cite{Kharzeev:2014xta}:
\begin{equation}
	\begin{aligned}\langle\bar{\psi}\psi(x_t)\rangle&=(1-\pi a^2P(x_t))\langle\bar{\psi}\psi\rangle_0\\&+\pi a^2P(x_t)\langle\bar{\psi}\psi\rangle_{E^{1+1}}.\end{aligned}
	\label{eq:total condensate}
\end{equation}
In \cite{Kharzeev:2014xta}, the authors showed that transverse size of string $a$ appears comparable to the lattice spacing. In our work, the parameter $a$ is used to reconcile the consistency between the model and lattice simulations \cite{Iritani:2015zwa}. For example, the transverse color flux distributions excited by different quark-antiquark color sources show slight differences. We take the results of the chiral condensate restoration ratio published in lattice simulations \cite{Iritani:2015zwa} as a reference and find that values of $a$ between $1.12a_{lat}$ and $1.25a_{lat}$ with lattice space $a_{lat}$ are reasonable. In this case, the transverse size of the color flux tube is very small, almost matching the lattice spacing.

Quantitatively calculating the impact of confining flux tube on chiral condensate has been the subject of numerous lattice studies \cite{Kaneko:2013jla,Iritani:2013apa,Iritani:2013rla,Iritani:2014fga,Iritani:2015zwa,Iritani:2016fvi}. It is worth noting that in the Ref \cite{Iritani:2015zwa}, the authors extended the confining string formation to two quarks, introducing the influence of a third quark. For specific details on the restoration of partial chiral symmetry, please refer to Refs. \cite{Kharzeev:2014xta,Kaneko:2013jla,Iritani:2013apa,Iritani:2013rla,Iritani:2014fga,Iritani:2015zwa,Iritani:2016fvi}.

\subsection{Entanglement entropy}
The above discussions are all based on the influence of background fields on vacuum chiral condensate. We now consider what specific changes should occur in the Hilbert space of vacuum condensate after the introduction of background fields. When there are no confining flux tubes present, the information of the vacuum chiral condensate is completely knowable, depending only on the model's coupling constant $g$ and some mathematical parameters. At this point, we can speculate that it is in a quantum pure state, as it satisfies the conditions of a pure state. However, when the background field is introduced, with the partial restoration of chiral symmetry, it can be understood as the energy condensed in the vacuum appearing as a``depression" in the transverse plane, which is caused by the confining flux tube \cite{Iritani:2016fvi}. Throughout the entire $x_t$ plane, the states space is decomposed into (i) the chiral condensate space ($A$) without the background field and (ii) the chiral condensate space ($B$) affected by the background field. The physical states inside the region $A$ ``probed" as the chiral condensate states without background color field in Hilbert space $\mathcal{H}_A$, and states is affected by background field $\mathcal{H}_B$.

The composite system in $A\cap B$ (the entire condensate) is then described by the vector $\Psi_{AB}$ in the space $\mathcal{H}_{A}\otimes\mathcal{H}_{B}$ that is a tensor product of the two space:
\begin{equation}
	|\Psi_{AB}\rangle=\sum_{i,j}c_{ij}|\varphi_i^A\rangle\otimes|\varphi_j^B\rangle,
	\label{eq:state}
\end{equation}
where $c_{i,j}$ are the elements of the matrix $C$ that has the dimension $n_A \times n_B$. If the sum (\ref{eq:state}) contains only one term, then the state $	|\Psi_{AB}\rangle$ is separable, or a product state. Otherwise the state $|\Psi_{AB}\rangle$ is entangled. The corresponding density matrix is $\rho_{AB}=|\Psi_{AB}\rangle\langle\Psi_{AB}|$.

The Schmidt decomposition theorem \cite{Schmidt:1907pp,peres1995quantum} allows the pure state $|\Psi_{AB}\rangle$ to be expanded as a single sum
\begin{equation}
	|\Psi_{AB}\rangle=\sum_n\alpha_n|\Psi_n^A\rangle|\Psi_n^B\rangle 
	\label{eq:Schmidt}
\end{equation}
for a suitably chosen orthonormal sets of $|\Psi_n^A\rangle$ and $|\Psi_n^B\rangle$ localized in the domains $A$ and $B$, respectively, where $\alpha_n$ are positive and real numbers that are the square roots of the eigenvalues of matrix $CC^\dagger$. Thus one can write down the density matrix $\rho_{AB}$ of the mixed state probed in region $A$ as
\begin{equation}
	\hat{\rho}_A=\mathrm{tr}_B\rho_{AB}=\sum_n\alpha_n^2|\Psi_n^A\rangle\langle\Psi_n^A|,
	\label{eq: rhoA}
\end{equation}
where $\alpha_n^2\equiv P_n$ is the probability of a state $|\Psi_n^A\rangle$. Now let's discuss the states $|\Psi_n^A\rangle$ and $|\Psi_n^B\rangle$ mentioned above. In fact, we do not have a straightforward method to construct the specific form of the wave function corresponding to a state after the Schmidt decomposition. However, under the assumptions we made earlier, we can acknowledge that the two states in Eq. (\ref{eq:Schmidt}) respectively represent the information of the two subspaces obtained by the direct product decomposition of the Hilbert space. Based on this, we can define the von Neumann entropy
\begin{equation}
	S_{EE}=-\sum_nP_n\ln P_n.
	\label{eq:EE}
\end{equation}

In our model, the probabilities $P_n$ only have two terms: $P_1=\pi a^2P(x_t)$ and $P_2=1-\pi a^2P(x_t)$ correspond to the probabilities of finding or not finding the flux at a given transverse position $x_t$. We need to emphasize that the von Neumann entropy (\ref{eq:EE}) here represents the Shannon entropy of the probability distribution ($P_n$) in information theory. In fact, the rigorous computation of entanglement entropy should be attributed to the exact density matrix trace of the system's quantum state construction. Here, we adopt the concept of information entropy to vividly illustrate to the reader that introducing external color sources in the vacuum will lead to changes in the information entropy corresponding to the original vacuum condensate, the extent of which depends on the estimation of von Neumann entropy. The introduced classical probability distribution aims to simulate the influence of the transverse distribution of the color field on chiral condensation. In summary, the information entropy presented here maps how the introduction of the color field affects the image of vacuum quark condensation, and achieving an accurate quantum entanglement description requires a significant amount of work.

\section{Numerical results and discussions}
\label{sec:dissc}
Before presenting our numerical results, we now briefly introduce how the background field $E_{3+1}(x_t)$, which is used to obtain $E_{1+1}$ and $P(x_t)$, is chosen. We utilize lattice calculation results from \cite{Cea:2012qw,Baker:2019gsi} as inputs for the distribution of color flux tubes, emphasizing that the distribution of color flux tubes is essentially independent of the longitudinal position of the probe, as explained in \cite{Cea:2012qw,Kharzeev:2014xta,Baker:2019gsi,Cardaci:2010tb,Cea:1995zt}. The measured color electric field as a function of the transverse coordinate can be well described by the following parameterization \cite{Clem:1975jr}
\begin{equation}
	E_{3+1}(x_t)=\frac\varphi{2\pi}\frac{\mu^2}\alpha\frac{K_0[(\mu^2x_t^2+\alpha^2)^{1/2}]}{K_1[\alpha]},
	\label{eq:E_lattice}
\end{equation}
where $\varphi$, $\alpha$ and $\mu$ are fitting parameters. The parameter $\varphi$ represents the external flux, and $\mu$ is the inverse of the London penetration depth $\lambda$. When the London penetration depth is much larger than the coherence length of the magnetic monopole condensate, i.e., when the Ginzburg-Landau parameter is much greater than 1, the chromomagnetic field is dual to the type II superconductor (see \cite{Cea:2012qw,Clem:1975jr}). Additionally, $\alpha$ represents the ratio of the vortex core radius to the London penetration depth, characterizing the scale of the vortex core. Eq. (\ref{eq:E_lattice}) is a simple analytical expression derived from Ampère's law and the Ginzburg-Landau equation. In particular, it can be simplified to the London model outside the vortex core in the presence of a transverse magnetic field distribution. It reflects the compatibility of color flux tubes with dual superconducting theories.

\begin{figure*}[htbp]
	\begin{center}
			\includegraphics[width=0.45\linewidth]{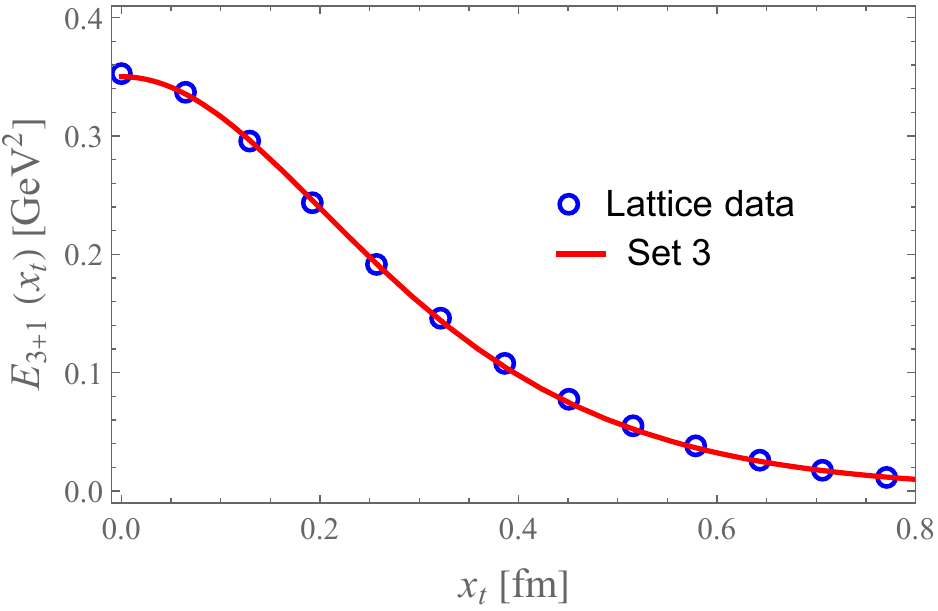}
		\includegraphics[width=0.45\linewidth]{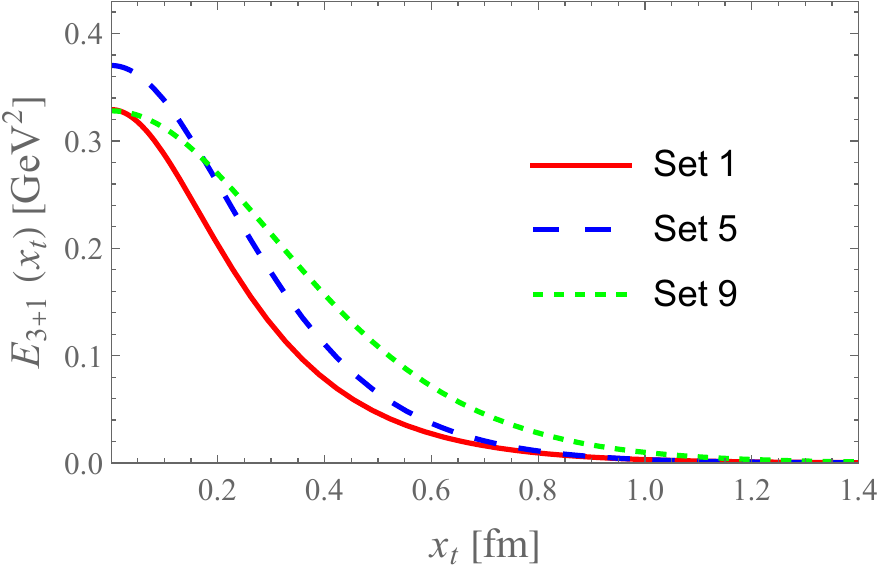}
		\includegraphics[width=0.45\linewidth]{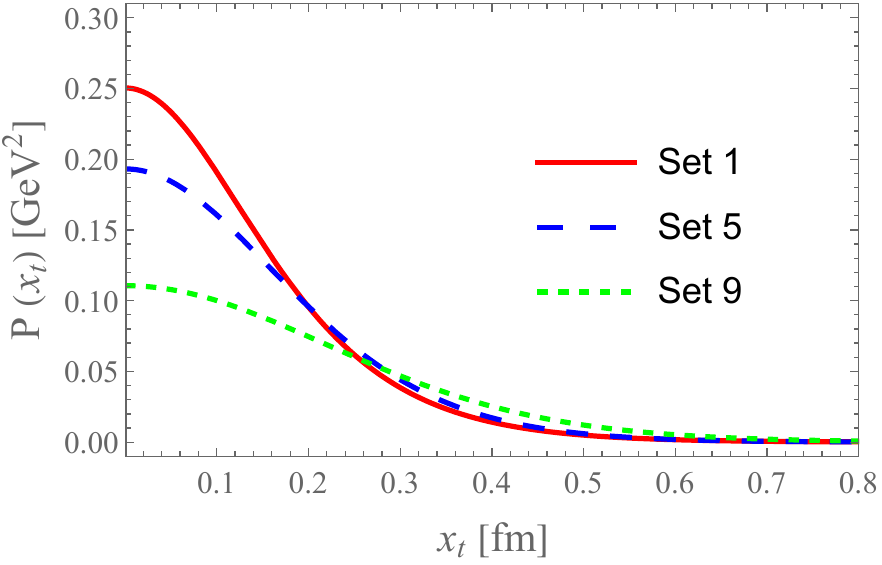}
		\includegraphics[width=0.45\linewidth]{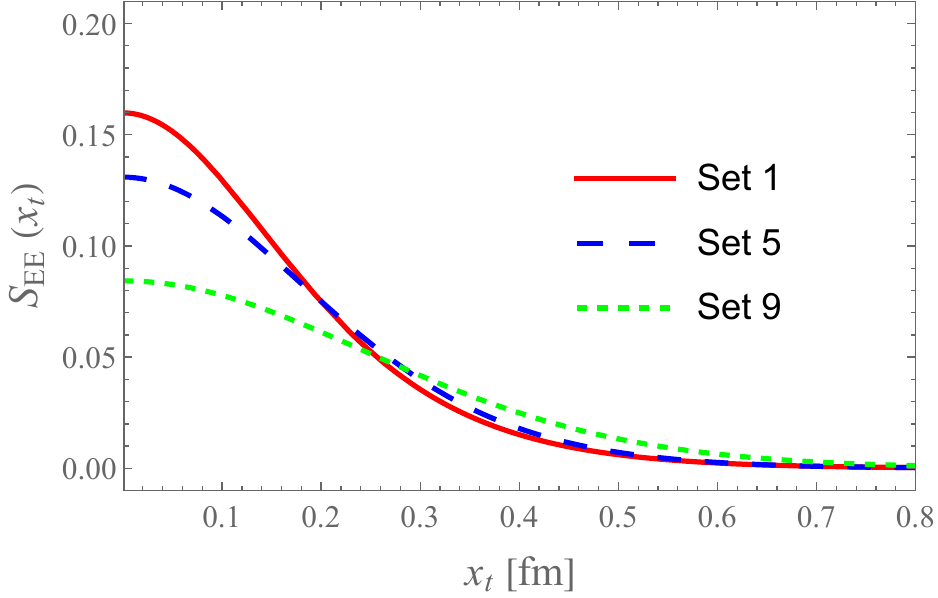}
	\end{center}
	\caption{Comparison of lattice simulation (blue circle) with the parameterization (top left). The transverse distribution of the color electric field existing between two static color charges (top right), along with the corresponding transverse probability (bottom left) and entanglement entropy (bottom right) distributions. The 1st, 3rd, 5th, and 9th parameter sets from Table 3 in Ref. \cite{Baker:2019gsi} are chosen.}
	\label{fig:distribution}
\end{figure*}

In this work, we adopt the lattice results from Ref. \cite{Baker:2019gsi}, which include the distribution of color flux tubes obtained as input from distances between different color charges. The relevant parameters are displayed in Ref. \cite{Baker:2019gsi} (See Table 3 in the reference). The lattice coupling constant $\beta=2N_c/g^2$ with color number $N_c$.
	To demonstrate the feasibility of parameterizing Eq. (\ref{eq:E_lattice}), we compare the lattice simulation results of $E_{3+1}(x_t)$ at the color source distance $d=0.51$ fm \cite{Baker:2019gsi} with the parameterization (see the top left of Figure \ref{fig:distribution}). Firstly, we chose the parameter sets of Set 1, 5, and 9 ($d=0.37$, 0.69 and 1.06 fm) to plot the profile of the color electric field $E_{3+1}(x_t)$ as a function of the transverse coordinate, formed by color charges at different distances. It can be seen that the profiles at different $d$ are distinct. Additionally, we provide the plots of $P(x_t)$ and $S_{EE}(x_t)$ by Eqs. (\ref{eq:3+1=1+1}, \ref{eq:EE}), see Figure \ref{fig:distribution}. We observe that the color field distribution, the probability distribution, and the entanglement entropy distribution all resemble ``Gaussian" distributions. This is based on the results of the lattice simulation and fitting using the Clem parameterized model \cite{Clem:1975jr}. This indicates that the rise and fall of a color flux tube excited between two static color charges in the transverse plane also exhibits Gaussian-like behavior. Furthermore, it suggests that the probability of finding the tube is highest at the origin of the polar coordinates (assuming the effective radius of the flux, $a$, is known). 

We now investigate the variation of the chiral condensate with respect to the background field. Initially, we provide the square root of the string tension corresponding to the static color charge from the position separation by Eq. (\ref{eq:3+1=1+1}). Following this, we integrate the chiral condensate under the influence of the background field, which represents the total condensate contribution. We anticipate that the total condensate value will differ for various color field backgrounds. For convenience, we include the integral of the vacuum condensate over the entire space, denoted as $\langle \bar{\psi}\psi\rangle^{int}=\frac1g\int d^2x_t \pi a^2 P(x_t) \langle \bar{\psi}\psi\rangle_{E^{1+1}}-C_0$ with $C_0=\frac1g\int d^2 x_t \langle \bar{\psi}\psi\rangle_0$. Finally, we examine the integration results for the entanglement entropy distribution. All the numerical results are presented in Table \ref{tab:results}, corresponding to the parameter selections listed in Table 3 in Ref. \cite{Baker:2019gsi}, respectively. We emphasize that all numerical integration rests on the rotational symmetry of the field distribution.

\begin{table}[htbp]
	\caption{The numerical results corresponding to different sets of parameters in \cite{Baker:2019gsi}, including $\sigma_E$, $\langle \bar{\psi}\psi\rangle^{int}$, and $S_{EE}^{int}$. }
		\begin{tabular}{l|l|l|l|l|l}
			\hline
			Sets & $\beta$ & $d$ (fm) & $\sqrt{\sigma_E}$ (GeV)& $\langle \bar{\psi}\psi\rangle^{int}$ (fm$^2$) & $S_{EE}^{int}$ (GeV$^{-2}$) \\ \hline
			1    & 6.475   & 0.37     & 0.4659(7)      & 1.9657(2)                             & 0.8361(4)                 \\
			2    & 6.333   & 0.45     & 0.5038(49)     & 1.973(13)                             & 0.8478(22)                \\
			3    & 6.240   & 0.51     & 0.5428(23)    & 1.9814(6)                             & 0.8469(11)                \\
			4    & 6.500   & 0.54     & 0.5681(35)     & 1.9929(10)                            & 0.8567(18)                \\
			5    & 6.539   & 0.69     & 0.597(17)     & 2.0018(48)                            & 0.8543(82)                \\
			6    & 6.370   & 0.85     & 0.638(77)    & 2.011(23)                             & 0.8962(369)               \\
			7    & 6.299   & 0.94     & 0.63(14)    & 2.006(42)                             & 0.8938(703)               \\
			8    & 6.240   & 1.02     & 0.75(12)    & 2.044(37)                             & 0.9563(469)               \\
			9    & 6.218   & 1.06     & 0.70(18)   & 2.026(56)                             & 0.9294(804)               \\ \hline
		\end{tabular}%
	\label{tab:results}
\end{table}

\begin{figure*}[htpb]
	\begin{center}
		\includegraphics[width=0.32\linewidth]{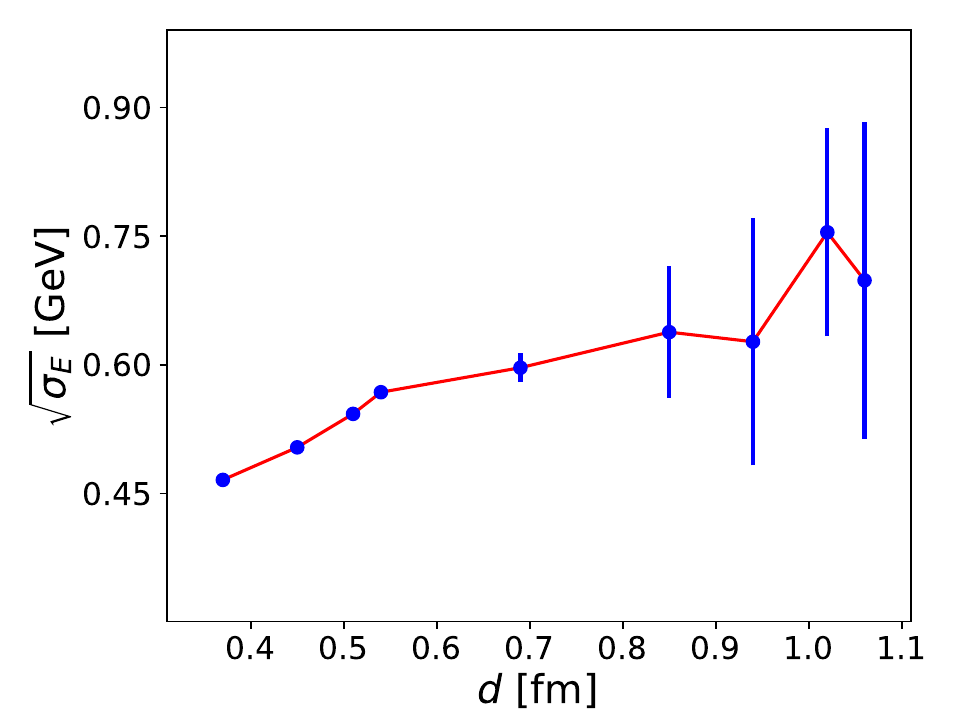}
		\includegraphics[width=0.32\linewidth]{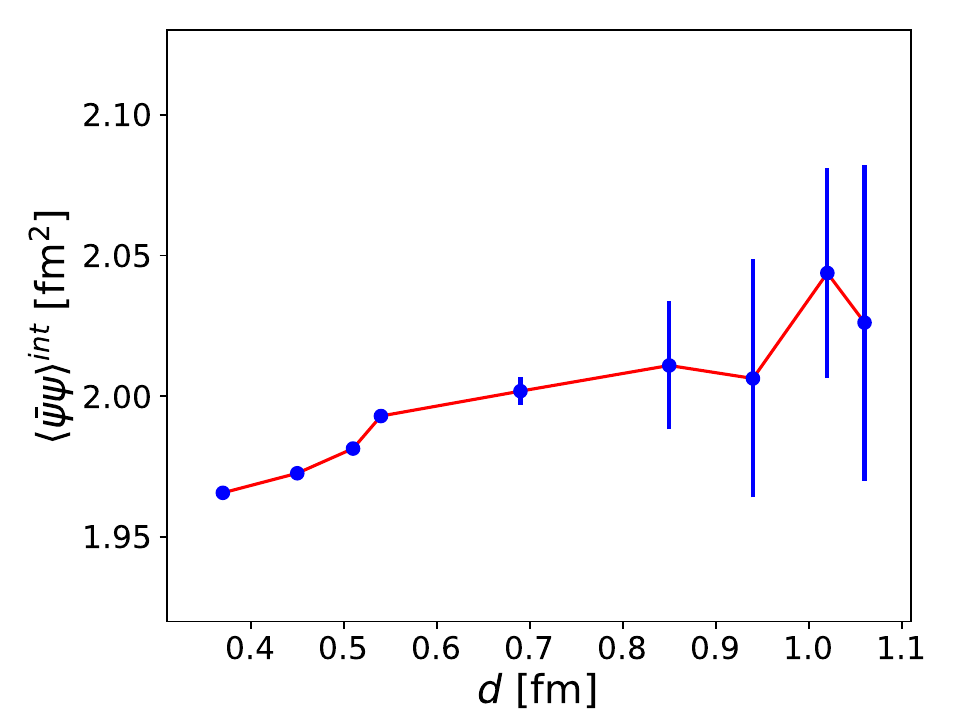}
		\includegraphics[width=0.32\linewidth]{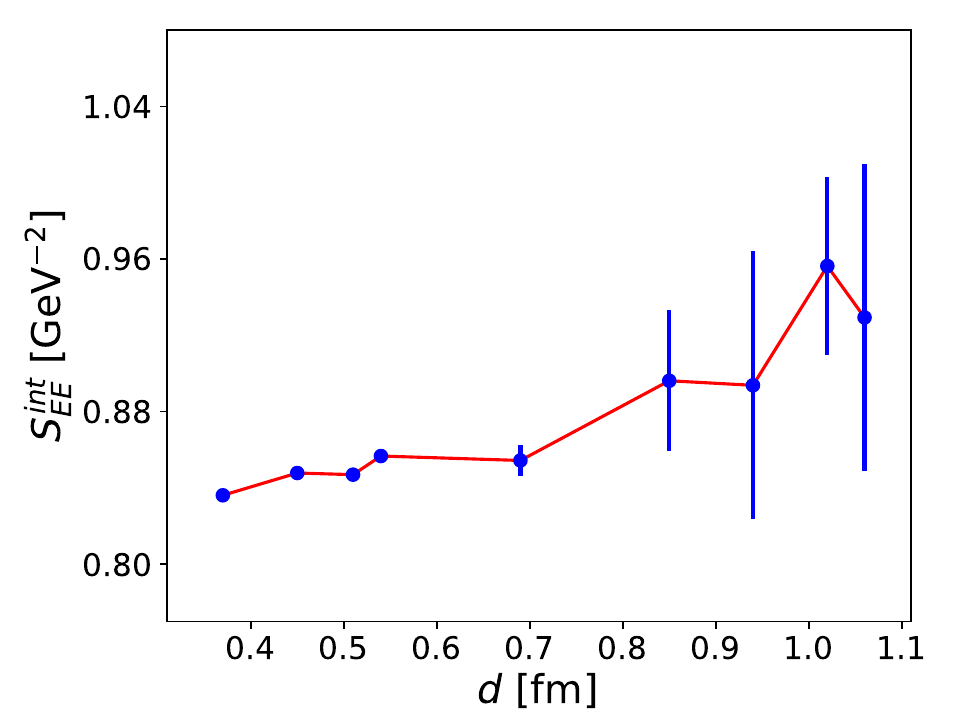}
	\end{center}
	\caption{Numerical results of string tensions, integrated chiral condensates, and integrated entanglement entropy at different color-charge distances $d$, with uncertainties from a Clem parameterized model \cite{Clem:1975jr} fit to the lattice simulation \cite{Baker:2019gsi}.}
	\label{fig:results}
\end{figure*}

Our numerical results are also displayed in Figure \ref{fig:results}. The uncertainties arise from the calculation of the uncertainty covariance matrix for the parameters in \cite{Baker:2019gsi}. It can be observed that the results for all three sets of physical quantities increase as the color charge distance $d$ increases. However, it is important to clarify the presence of large uncertainties in the current results.

The variations of physical quantities with $d$ in Figure \ref{fig:results} are clearly attributed to the distribution of color flux tubes between positive and negative color charges at different distances. The string tension defines the energy contained in a unit length of the color flux tube, and its increase with $d$ is a result obtained from lattice simulations. Based on this premise, the two remaining physical quantities are both associated with the probability distribution $P(x_t)$ introduced by the conversion of $E_{3+1}$ to $E_{1+1}$, and after integration over the transverse space, they both degenerate to the $E_{1+1}$ dependence, similar to the cosine oscillatory behavior of chiral condensate under the influence of the background field (\ref{eq:chiral condensate}). As the separation of color sources increases, the thickness of the flux may be scaled by a logarithmic function with respect to its length \cite{Hasenfratz:1980ue,Luscher:1980iy}, and this information is encoded in the energy per unit length, i.e., the string tension. Such behavior has indeed been studied in quenched lattice QCD calculations \cite{Cardoso:2013lla} as well as in finite temperature \cite{Bakry:2010sp}. This directly affects the change in vacuum chiral condensate, and precisely reduces the amount of condensate, which corresponds to a partial chiral symmetry restoration \cite{Iritani:2015zwa}.

The integrated value of entanglement entropy also increases with the color source separation $d$, which can be interpreted as an increase in the effect of string tension on the chiral condensate when the color source flux thickness is increased. In other words, the vacuum chiral condensate decreases, the degrees of freedom in $A$ space increase, and the entanglement between $A$ and $B$ (the complement of $A$) is increased. We can conjecture that this process will be similar to the entanglement entropy change process of two quantum subsystems into which other similar pure state systems have been separated. When the chiral condensate of the whole space completely disappears after the influence of the flux with infinite thickness, then the chiral symmetry is completely recovered and the entanglement entropy goes to zero.

\section{Summary and outlook}
\label{sec:summary}
In this work, we mainly provide three discussions: (i) We reviewed the method of calculating the chiral condensate using the Schwinger model and introduced the influence of the vacuum color flux tube on the vacuum chiral condensate. (ii) We utilized the color electric field distribution obtained from recent lattice calculations \cite{Baker:2019gsi}, corresponding to different color source separations. We found that as the color source separation increases, the corresponding thickness of the color flux tube increases, which affects the transverse distribution shape of the color field, thereby influencing the string tension. (iii) The string tension increases with the increase of color source separation and affects the amount of chiral condensate recovery. We observed that the recovery amount increases with the increase of color source separation. Additionally, we hypothesize the influence of color sources as the basis for the direct product decomposition of the entire system's Hilbert space and provide the change in the system's entanglement entropy. We found that the entanglement entropy also increases with the increase of color source separation. The entanglement entropy corresponding to the chiral condensate may be induced by the entanglement between two color sources, and the information encoded in the entanglement among two color sources may be discovered from the observation of the chiral condensate.

The current numerical results depend on the uncertainties of the parameters corresponding to the color electric field distribution obtained from lattice calculations \cite{Baker:2019gsi}. It can be observed that parameters at larger separations of color charges exist higher fitting uncertainties. This is related to the lattice configurations in lattice QCD calculations, where it is difficult to satisfy the lattice spacing limit at the edges of the lattice. Meanwhile, it is also interesting to investigate the effects of introducing the third color source on the color flux tube \cite{Iritani:2015zwa,Iritani:2016fvi}. It is worth considering how the influence caused by the third color source is reflected in the entanglement entropy. Additionally, vacuum polarization of oppositely moving fermion pairs and string breaking-induced vacuum fluctuations also increase the entanglement entropy between color sources and the vacuum fermion condensate. This has been discussed in recent works in detail \cite{Florio:2023dke,Ikeda:2023zil}. Whether there are  connections between our present study and the recent works \cite{Florio:2023dke,Ikeda:2023zil} could be a direction for us to explore in the future.

\begin{acknowledgments}

This work is supported by the Strategic Priority Research Program of Chinese Academy of Sciences under the Grant NO. XDB34030301.
\end{acknowledgments}

\bibliographystyle{apsrev4-1}
\bibliography{refs}

\begin{thebibliography}{52}%
\makeatletter
\providecommand \@ifxundefined [1]{%
 \@ifx{#1\undefined}
}%
\providecommand \@ifnum [1]{%
 \ifnum #1\expandafter \@firstoftwo
 \else \expandafter \@secondoftwo
 \fi
}%
\providecommand \@ifx [1]{%
 \ifx #1\expandafter \@firstoftwo
 \else \expandafter \@secondoftwo
 \fi
}%
\providecommand \natexlab [1]{#1}%
\providecommand \enquote  [1]{``#1''}%
\providecommand \bibnamefont  [1]{#1}%
\providecommand \bibfnamefont [1]{#1}%
\providecommand \citenamefont [1]{#1}%
\providecommand \href@noop [0]{\@secondoftwo}%
\providecommand \href [0]{\begingroup \@sanitize@url \@href}%
\providecommand \@href[1]{\@@startlink{#1}\@@href}%
\providecommand \@@href[1]{\endgroup#1\@@endlink}%
\providecommand \@sanitize@url [0]{\catcode `\\12\catcode `\$12\catcode
  `\&12\catcode `\#12\catcode `\^12\catcode `\_12\catcode `\%12\relax}%
\providecommand \@@startlink[1]{}%
\providecommand \@@endlink[0]{}%
\providecommand \url  [0]{\begingroup\@sanitize@url \@url }%
\providecommand \@url [1]{\endgroup\@href {#1}{\urlprefix }}%
\providecommand \urlprefix  [0]{URL }%
\providecommand \Eprint [0]{\href }%
\providecommand \doibase [0]{http://dx.doi.org/}%
\providecommand \selectlanguage [0]{\@gobble}%
\providecommand \bibinfo  [0]{\@secondoftwo}%
\providecommand \bibfield  [0]{\@secondoftwo}%
\providecommand \translation [1]{[#1]}%
\providecommand \BibitemOpen [0]{}%
\providecommand \bibitemStop [0]{}%
\providecommand \bibitemNoStop [0]{.\EOS\space}%
\providecommand \EOS [0]{\spacefactor3000\relax}%
\providecommand \BibitemShut  [1]{\csname bibitem#1\endcsname}%
\let\auto@bib@innerbib\@empty
\bibitem [{\citenamefont {Schwinger}(1962{\natexlab{a}})}]{Schwinger:1962tn}%
  \BibitemOpen
  \bibfield  {author} {\bibinfo {author} {\bibfnamefont {J.~S.}\ \bibnamefont
  {Schwinger}},\ }\href {\doibase 10.1103/PhysRev.125.397} {\bibfield
  {journal} {\bibinfo  {journal} {Phys. Rev.}\ }\textbf {\bibinfo {volume}
  {125}},\ \bibinfo {pages} {397} (\bibinfo {year}
  {1962}{\natexlab{a}})}\BibitemShut {NoStop}%
\bibitem [{\citenamefont {Schwinger}(1962{\natexlab{b}})}]{Schwinger:1962tp}%
  \BibitemOpen
  \bibfield  {author} {\bibinfo {author} {\bibfnamefont {J.~S.}\ \bibnamefont
  {Schwinger}},\ }\href {\doibase 10.1103/PhysRev.128.2425} {\bibfield
  {journal} {\bibinfo  {journal} {Phys. Rev.}\ }\textbf {\bibinfo {volume}
  {128}},\ \bibinfo {pages} {2425} (\bibinfo {year}
  {1962}{\natexlab{b}})}\BibitemShut {NoStop}%
\bibitem [{\citenamefont {Brown}(1963)}]{Brown:1963po}%
  \BibitemOpen
  \bibfield  {author} {\bibinfo {author} {\bibfnamefont {L.~S.}\ \bibnamefont
  {Brown}},\ }\href {\doibase 10.1007/BF02827786} {\bibfield  {journal}
  {\bibinfo  {journal} {Il Nuovo Cimento (1955-1965)}\ }\textbf {\bibinfo
  {volume} {29}},\ \bibinfo {pages} {617} (\bibinfo {year} {1963})}\BibitemShut
  {NoStop}%
\bibitem [{\citenamefont {Zumino}(1964)}]{Zumino:1964nja}%
  \BibitemOpen
  \bibfield  {author} {\bibinfo {author} {\bibfnamefont {B.}~\bibnamefont
  {Zumino}},\ }\href {\doibase 10.1016/0031-9163(64)90177-5} {\bibfield
  {journal} {\bibinfo  {journal} {Phys. Lett.}\ }\textbf {\bibinfo {volume}
  {10}},\ \bibinfo {pages} {224} (\bibinfo {year} {1964})}\BibitemShut
  {NoStop}%
\bibitem [{\citenamefont {Lowenstein}\ and\ \citenamefont
  {Swieca}(1971)}]{Lowenstein:1971fc}%
  \BibitemOpen
  \bibfield  {author} {\bibinfo {author} {\bibfnamefont {J.~H.}\ \bibnamefont
  {Lowenstein}}\ and\ \bibinfo {author} {\bibfnamefont {J.~A.}\ \bibnamefont
  {Swieca}},\ }\href {\doibase 10.1016/0003-4916(71)90246-6} {\bibfield
  {journal} {\bibinfo  {journal} {Annals Phys.}\ }\textbf {\bibinfo {volume}
  {68}},\ \bibinfo {pages} {172} (\bibinfo {year} {1971})}\BibitemShut
  {NoStop}%
\bibitem [{\citenamefont {Casher}\ \emph {et~al.}(1974)\citenamefont {Casher},
  \citenamefont {Kogut},\ and\ \citenamefont {Susskind}}]{Casher:1974vf}%
  \BibitemOpen
  \bibfield  {author} {\bibinfo {author} {\bibfnamefont {A.}~\bibnamefont
  {Casher}}, \bibinfo {author} {\bibfnamefont {J.~B.}\ \bibnamefont {Kogut}}, \
  and\ \bibinfo {author} {\bibfnamefont {L.}~\bibnamefont {Susskind}},\ }\href
  {\doibase 10.1103/PhysRevD.10.732} {\bibfield  {journal} {\bibinfo  {journal}
  {Phys. Rev. D}\ }\textbf {\bibinfo {volume} {10}},\ \bibinfo {pages} {732}
  (\bibinfo {year} {1974})}\BibitemShut {NoStop}%
\bibitem [{\citenamefont {Coleman}\ \emph {et~al.}(1975)\citenamefont
  {Coleman}, \citenamefont {Jackiw},\ and\ \citenamefont
  {Susskind}}]{Coleman:1975pw}%
  \BibitemOpen
  \bibfield  {author} {\bibinfo {author} {\bibfnamefont {S.~R.}\ \bibnamefont
  {Coleman}}, \bibinfo {author} {\bibfnamefont {R.}~\bibnamefont {Jackiw}}, \
  and\ \bibinfo {author} {\bibfnamefont {L.}~\bibnamefont {Susskind}},\ }\href
  {\doibase 10.1016/0003-4916(75)90212-2} {\bibfield  {journal} {\bibinfo
  {journal} {Annals Phys.}\ }\textbf {\bibinfo {volume} {93}},\ \bibinfo
  {pages} {267} (\bibinfo {year} {1975})}\BibitemShut {NoStop}%
\bibitem [{\citenamefont {Kharzeev}\ and\ \citenamefont
  {Loshaj}(2014)}]{Kharzeev:2014xta}%
  \BibitemOpen
  \bibfield  {author} {\bibinfo {author} {\bibfnamefont {D.~E.}\ \bibnamefont
  {Kharzeev}}\ and\ \bibinfo {author} {\bibfnamefont {F.}~\bibnamefont
  {Loshaj}},\ }\href {\doibase 10.1103/PhysRevD.90.037501} {\bibfield
  {journal} {\bibinfo  {journal} {Phys. Rev. D}\ }\textbf {\bibinfo {volume}
  {90}},\ \bibinfo {pages} {037501} (\bibinfo {year} {2014})},\ \Eprint
  {http://arxiv.org/abs/1404.7746} {arXiv:1404.7746 [hep-ph]} \BibitemShut
  {NoStop}%
\bibitem [{\citenamefont {Tomiya}(2022)}]{Tomiya:2022chr}%
  \BibitemOpen
  \bibfield  {author} {\bibinfo {author} {\bibfnamefont {A.}~\bibnamefont
  {Tomiya}},\ }\href@noop {} {\  (\bibinfo {year} {2022})},\ \Eprint
  {http://arxiv.org/abs/2205.08860} {arXiv:2205.08860 [hep-lat]} \BibitemShut
  {NoStop}%
\bibitem [{\citenamefont {Florio}\ \emph {et~al.}(2023)\citenamefont {Florio},
  \citenamefont {Frenklakh}, \citenamefont {Ikeda}, \citenamefont {Kharzeev},
  \citenamefont {Korepin}, \citenamefont {Shi},\ and\ \citenamefont
  {Yu}}]{Florio:2023dke}%
  \BibitemOpen
  \bibfield  {author} {\bibinfo {author} {\bibfnamefont {A.}~\bibnamefont
  {Florio}}, \bibinfo {author} {\bibfnamefont {D.}~\bibnamefont {Frenklakh}},
  \bibinfo {author} {\bibfnamefont {K.}~\bibnamefont {Ikeda}}, \bibinfo
  {author} {\bibfnamefont {D.}~\bibnamefont {Kharzeev}}, \bibinfo {author}
  {\bibfnamefont {V.}~\bibnamefont {Korepin}}, \bibinfo {author} {\bibfnamefont
  {S.}~\bibnamefont {Shi}}, \ and\ \bibinfo {author} {\bibfnamefont
  {K.}~\bibnamefont {Yu}},\ }\href {\doibase 10.1103/PhysRevLett.131.021902}
  {\bibfield  {journal} {\bibinfo  {journal} {Phys. Rev. Lett.}\ }\textbf
  {\bibinfo {volume} {131}},\ \bibinfo {pages} {021902} (\bibinfo {year}
  {2023})},\ \Eprint {http://arxiv.org/abs/2301.11991} {arXiv:2301.11991
  [hep-ph]} \BibitemShut {NoStop}%
\bibitem [{\citenamefont {Xiao}(2008)}]{Xiao:2008nr}%
  \BibitemOpen
  \bibfield  {author} {\bibinfo {author} {\bibfnamefont {B.-W.}\ \bibnamefont
  {Xiao}},\ }\href {\doibase 10.1016/j.physletb.2008.06.017} {\bibfield
  {journal} {\bibinfo  {journal} {Phys. Lett. B}\ }\textbf {\bibinfo {volume}
  {665}},\ \bibinfo {pages} {173} (\bibinfo {year} {2008})},\ \Eprint
  {http://arxiv.org/abs/0804.1343} {arXiv:0804.1343 [hep-th]} \BibitemShut
  {NoStop}%
\bibitem [{\citenamefont {Semenoff}\ and\ \citenamefont
  {Zarembo}(2011)}]{Semenoff:2011ng}%
  \BibitemOpen
  \bibfield  {author} {\bibinfo {author} {\bibfnamefont {G.~W.}\ \bibnamefont
  {Semenoff}}\ and\ \bibinfo {author} {\bibfnamefont {K.}~\bibnamefont
  {Zarembo}},\ }\href {\doibase 10.1103/PhysRevLett.107.171601} {\bibfield
  {journal} {\bibinfo  {journal} {Phys. Rev. Lett.}\ }\textbf {\bibinfo
  {volume} {107}},\ \bibinfo {pages} {171601} (\bibinfo {year} {2011})},\
  \Eprint {http://arxiv.org/abs/1109.2920} {arXiv:1109.2920 [hep-th]}
  \BibitemShut {NoStop}%
\bibitem [{\citenamefont {Chernicoff}\ \emph {et~al.}(2013)\citenamefont
  {Chernicoff}, \citenamefont {G\"uijosa},\ and\ \citenamefont
  {Pedraza}}]{Chernicoff:2013iga}%
  \BibitemOpen
  \bibfield  {author} {\bibinfo {author} {\bibfnamefont {M.}~\bibnamefont
  {Chernicoff}}, \bibinfo {author} {\bibfnamefont {A.}~\bibnamefont
  {G\"uijosa}}, \ and\ \bibinfo {author} {\bibfnamefont {J.~F.}\ \bibnamefont
  {Pedraza}},\ }\href {\doibase 10.1007/JHEP10(2013)211} {\bibfield  {journal}
  {\bibinfo  {journal} {JHEP}\ }\textbf {\bibinfo {volume} {10}},\ \bibinfo
  {pages} {211} (\bibinfo {year} {2013})},\ \Eprint
  {http://arxiv.org/abs/1308.3695} {arXiv:1308.3695 [hep-th]} \BibitemShut
  {NoStop}%
\bibitem [{\citenamefont {Jensen}\ and\ \citenamefont
  {Karch}(2013)}]{Jensen:2013ora}%
  \BibitemOpen
  \bibfield  {author} {\bibinfo {author} {\bibfnamefont {K.}~\bibnamefont
  {Jensen}}\ and\ \bibinfo {author} {\bibfnamefont {A.}~\bibnamefont {Karch}},\
  }\href {\doibase 10.1103/PhysRevLett.111.211602} {\bibfield  {journal}
  {\bibinfo  {journal} {Phys. Rev. Lett.}\ }\textbf {\bibinfo {volume} {111}},\
  \bibinfo {pages} {211602} (\bibinfo {year} {2013})},\ \Eprint
  {http://arxiv.org/abs/1307.1132} {arXiv:1307.1132 [hep-th]} \BibitemShut
  {NoStop}%
\bibitem [{\citenamefont {Lewkowycz}\ and\ \citenamefont
  {Maldacena}(2014)}]{Lewkowycz:2013laa}%
  \BibitemOpen
  \bibfield  {author} {\bibinfo {author} {\bibfnamefont {A.}~\bibnamefont
  {Lewkowycz}}\ and\ \bibinfo {author} {\bibfnamefont {J.}~\bibnamefont
  {Maldacena}},\ }\href {\doibase 10.1007/JHEP05(2014)025} {\bibfield
  {journal} {\bibinfo  {journal} {JHEP}\ }\textbf {\bibinfo {volume} {05}},\
  \bibinfo {pages} {025} (\bibinfo {year} {2014})},\ \Eprint
  {http://arxiv.org/abs/1312.5682} {arXiv:1312.5682 [hep-th]} \BibitemShut
  {NoStop}%
\bibitem [{\citenamefont {Sonner}(2013)}]{Sonner:2013mba}%
  \BibitemOpen
  \bibfield  {author} {\bibinfo {author} {\bibfnamefont {J.}~\bibnamefont
  {Sonner}},\ }\href {\doibase 10.1103/PhysRevLett.111.211603} {\bibfield
  {journal} {\bibinfo  {journal} {Phys. Rev. Lett.}\ }\textbf {\bibinfo
  {volume} {111}},\ \bibinfo {pages} {211603} (\bibinfo {year} {2013})},\
  \Eprint {http://arxiv.org/abs/1307.6850} {arXiv:1307.6850 [hep-th]}
  \BibitemShut {NoStop}%
\bibitem [{\citenamefont {Hubeny}\ and\ \citenamefont
  {Semenoff}(2014)}]{Hubeny:2014zna}%
  \BibitemOpen
  \bibfield  {author} {\bibinfo {author} {\bibfnamefont {V.~E.}\ \bibnamefont
  {Hubeny}}\ and\ \bibinfo {author} {\bibfnamefont {G.~W.}\ \bibnamefont
  {Semenoff}},\ }\href@noop {} {\  (\bibinfo {year} {2014})},\ \Eprint
  {http://arxiv.org/abs/1410.1172} {arXiv:1410.1172 [hep-th]} \BibitemShut
  {NoStop}%
\bibitem [{\citenamefont {Jensen}\ \emph {et~al.}(2014)\citenamefont {Jensen},
  \citenamefont {Karch},\ and\ \citenamefont {Robinson}}]{Jensen:2014bpa}%
  \BibitemOpen
  \bibfield  {author} {\bibinfo {author} {\bibfnamefont {K.}~\bibnamefont
  {Jensen}}, \bibinfo {author} {\bibfnamefont {A.}~\bibnamefont {Karch}}, \
  and\ \bibinfo {author} {\bibfnamefont {B.}~\bibnamefont {Robinson}},\ }\href
  {\doibase 10.1103/PhysRevD.90.064019} {\bibfield  {journal} {\bibinfo
  {journal} {Phys. Rev. D}\ }\textbf {\bibinfo {volume} {90}},\ \bibinfo
  {pages} {064019} (\bibinfo {year} {2014})},\ \Eprint
  {http://arxiv.org/abs/1405.2065} {arXiv:1405.2065 [hep-th]} \BibitemShut
  {NoStop}%
\bibitem [{\citenamefont {Ghodrati}(2015)}]{Ghodrati:2015rta}%
  \BibitemOpen
  \bibfield  {author} {\bibinfo {author} {\bibfnamefont {M.}~\bibnamefont
  {Ghodrati}},\ }\href {\doibase 10.1103/PhysRevD.92.065015} {\bibfield
  {journal} {\bibinfo  {journal} {Phys. Rev. D}\ }\textbf {\bibinfo {volume}
  {92}},\ \bibinfo {pages} {065015} (\bibinfo {year} {2015})},\ \Eprint
  {http://arxiv.org/abs/1506.08557} {arXiv:1506.08557 [hep-th]} \BibitemShut
  {NoStop}%
\bibitem [{\citenamefont {Grieninger}\ \emph
  {et~al.}(2023{\natexlab{a}})\citenamefont {Grieninger}, \citenamefont
  {Kharzeev},\ and\ \citenamefont {Zahed}}]{Grieninger:2023ehb}%
  \BibitemOpen
  \bibfield  {author} {\bibinfo {author} {\bibfnamefont {S.}~\bibnamefont
  {Grieninger}}, \bibinfo {author} {\bibfnamefont {D.~E.}\ \bibnamefont
  {Kharzeev}}, \ and\ \bibinfo {author} {\bibfnamefont {I.}~\bibnamefont
  {Zahed}},\ }\href {\doibase 10.1103/PhysRevD.108.086030} {\bibfield
  {journal} {\bibinfo  {journal} {Phys. Rev. D}\ }\textbf {\bibinfo {volume}
  {108}},\ \bibinfo {pages} {086030} (\bibinfo {year} {2023}{\natexlab{a}})},\
  \Eprint {http://arxiv.org/abs/2305.07121} {arXiv:2305.07121 [hep-th]}
  \BibitemShut {NoStop}%
\bibitem [{\citenamefont {Grieninger}\ \emph
  {et~al.}(2023{\natexlab{b}})\citenamefont {Grieninger}, \citenamefont
  {Kharzeev},\ and\ \citenamefont {Zahed}}]{Grieninger:2023pyb}%
  \BibitemOpen
  \bibfield  {author} {\bibinfo {author} {\bibfnamefont {S.}~\bibnamefont
  {Grieninger}}, \bibinfo {author} {\bibfnamefont {D.~E.}\ \bibnamefont
  {Kharzeev}}, \ and\ \bibinfo {author} {\bibfnamefont {I.}~\bibnamefont
  {Zahed}},\ }\href {\doibase 10.1103/PhysRevD.108.126014} {\bibfield
  {journal} {\bibinfo  {journal} {Phys. Rev. D}\ }\textbf {\bibinfo {volume}
  {108}},\ \bibinfo {pages} {126014} (\bibinfo {year} {2023}{\natexlab{b}})},\
  \Eprint {http://arxiv.org/abs/2310.12042} {arXiv:2310.12042 [hep-th]}
  \BibitemShut {NoStop}%
\bibitem [{\citenamefont {Ikeda}\ \emph {et~al.}(2023)\citenamefont {Ikeda},
  \citenamefont {Kharzeev}, \citenamefont {Meyer},\ and\ \citenamefont
  {Shi}}]{Ikeda:2023zil}%
  \BibitemOpen
  \bibfield  {author} {\bibinfo {author} {\bibfnamefont {K.}~\bibnamefont
  {Ikeda}}, \bibinfo {author} {\bibfnamefont {D.~E.}\ \bibnamefont {Kharzeev}},
  \bibinfo {author} {\bibfnamefont {R.}~\bibnamefont {Meyer}}, \ and\ \bibinfo
  {author} {\bibfnamefont {S.}~\bibnamefont {Shi}},\ }\href {\doibase
  10.1103/PhysRevD.108.L091501} {\bibfield  {journal} {\bibinfo  {journal}
  {Phys. Rev. D}\ }\textbf {\bibinfo {volume} {108}},\ \bibinfo {pages}
  {L091501} (\bibinfo {year} {2023})},\ \Eprint
  {http://arxiv.org/abs/2305.00996} {arXiv:2305.00996 [hep-ph]} \BibitemShut
  {NoStop}%
\bibitem [{\citenamefont {Nambu}\ and\ \citenamefont
  {Jona-Lasinio}(1961{\natexlab{a}})}]{Nambu:1961tp}%
  \BibitemOpen
  \bibfield  {author} {\bibinfo {author} {\bibfnamefont {Y.}~\bibnamefont
  {Nambu}}\ and\ \bibinfo {author} {\bibfnamefont {G.}~\bibnamefont
  {Jona-Lasinio}},\ }\href {\doibase 10.1103/PhysRev.122.345} {\bibfield
  {journal} {\bibinfo  {journal} {Phys. Rev.}\ }\textbf {\bibinfo {volume}
  {122}},\ \bibinfo {pages} {345} (\bibinfo {year}
  {1961}{\natexlab{a}})}\BibitemShut {NoStop}%
\bibitem [{\citenamefont {Nambu}\ and\ \citenamefont
  {Jona-Lasinio}(1961{\natexlab{b}})}]{Nambu:1961fr}%
  \BibitemOpen
  \bibfield  {author} {\bibinfo {author} {\bibfnamefont {Y.}~\bibnamefont
  {Nambu}}\ and\ \bibinfo {author} {\bibfnamefont {G.}~\bibnamefont
  {Jona-Lasinio}},\ }\href {\doibase 10.1103/PhysRev.124.246} {\bibfield
  {journal} {\bibinfo  {journal} {Phys. Rev.}\ }\textbf {\bibinfo {volume}
  {124}},\ \bibinfo {pages} {246} (\bibinfo {year}
  {1961}{\natexlab{b}})}\BibitemShut {NoStop}%
\bibitem [{\citenamefont {Hamer}\ \emph {et~al.}(1982)\citenamefont {Hamer},
  \citenamefont {Kogut}, \citenamefont {Crewther},\ and\ \citenamefont
  {Mazzolini}}]{Hamer:1982mx}%
  \BibitemOpen
  \bibfield  {author} {\bibinfo {author} {\bibfnamefont {C.~J.}\ \bibnamefont
  {Hamer}}, \bibinfo {author} {\bibfnamefont {J.~B.}\ \bibnamefont {Kogut}},
  \bibinfo {author} {\bibfnamefont {D.~P.}\ \bibnamefont {Crewther}}, \ and\
  \bibinfo {author} {\bibfnamefont {M.~M.}\ \bibnamefont {Mazzolini}},\ }\href
  {\doibase 10.1016/0550-3213(82)90229-2} {\bibfield  {journal} {\bibinfo
  {journal} {Nucl. Phys. B}\ }\textbf {\bibinfo {volume} {208}},\ \bibinfo
  {pages} {413} (\bibinfo {year} {1982})}\BibitemShut {NoStop}%
\bibitem [{\citenamefont {Singh}\ \emph {et~al.}(1993)\citenamefont {Singh},
  \citenamefont {Browne},\ and\ \citenamefont {Haymaker}}]{Singh:1993jj}%
  \BibitemOpen
  \bibfield  {author} {\bibinfo {author} {\bibfnamefont {V.}~\bibnamefont
  {Singh}}, \bibinfo {author} {\bibfnamefont {D.~A.}\ \bibnamefont {Browne}}, \
  and\ \bibinfo {author} {\bibfnamefont {R.~W.}\ \bibnamefont {Haymaker}},\
  }\href {\doibase 10.1016/0370-2693(93)91146-E} {\bibfield  {journal}
  {\bibinfo  {journal} {Phys. Lett. B}\ }\textbf {\bibinfo {volume} {306}},\
  \bibinfo {pages} {115} (\bibinfo {year} {1993})},\ \Eprint
  {http://arxiv.org/abs/hep-lat/9301004} {arXiv:hep-lat/9301004} \BibitemShut
  {NoStop}%
\bibitem [{\citenamefont {Schilling}\ \emph {et~al.}(1999)\citenamefont
  {Schilling}, \citenamefont {Bali},\ and\ \citenamefont
  {Schlichter}}]{Schilling:1998gz}%
  \BibitemOpen
  \bibfield  {author} {\bibinfo {author} {\bibfnamefont {K.}~\bibnamefont
  {Schilling}}, \bibinfo {author} {\bibfnamefont {G.~S.}\ \bibnamefont {Bali}},
  \ and\ \bibinfo {author} {\bibfnamefont {C.}~\bibnamefont {Schlichter}},\
  }\href {\doibase 10.1016/S0920-5632(99)85160-3} {\bibfield  {journal}
  {\bibinfo  {journal} {Nucl. Phys. B Proc. Suppl.}\ }\textbf {\bibinfo
  {volume} {73}},\ \bibinfo {pages} {638} (\bibinfo {year} {1999})},\ \Eprint
  {http://arxiv.org/abs/hep-lat/9809039} {arXiv:hep-lat/9809039} \BibitemShut
  {NoStop}%
\bibitem [{\citenamefont {Chernodub}\ \emph {et~al.}(2001)\citenamefont
  {Chernodub}, \citenamefont {Gubarev}, \citenamefont {Polikarpov},\ and\
  \citenamefont {Zakharov}}]{Chernodub:2000rg}%
  \BibitemOpen
  \bibfield  {author} {\bibinfo {author} {\bibfnamefont {M.~N.}\ \bibnamefont
  {Chernodub}}, \bibinfo {author} {\bibfnamefont {F.~V.}\ \bibnamefont
  {Gubarev}}, \bibinfo {author} {\bibfnamefont {M.~I.}\ \bibnamefont
  {Polikarpov}}, \ and\ \bibinfo {author} {\bibfnamefont {V.~I.}\ \bibnamefont
  {Zakharov}},\ }\href {\doibase 10.1016/S0550-3213(01)00011-6} {\bibfield
  {journal} {\bibinfo  {journal} {Nucl. Phys. B}\ }\textbf {\bibinfo {volume}
  {600}},\ \bibinfo {pages} {163} (\bibinfo {year} {2001})},\ \Eprint
  {http://arxiv.org/abs/hep-th/0010265} {arXiv:hep-th/0010265} \BibitemShut
  {NoStop}%
\bibitem [{\citenamefont {Chernodub}\ \emph {et~al.}(2005)\citenamefont
  {Chernodub}, \citenamefont {Ishiguro}, \citenamefont {Mori}, \citenamefont
  {Nakamura}, \citenamefont {Polikarpov}, \citenamefont {Sekido}, \citenamefont
  {Suzuki},\ and\ \citenamefont {Zakharov}}]{Chernodub:2005gz}%
  \BibitemOpen
  \bibfield  {author} {\bibinfo {author} {\bibfnamefont {M.~N.}\ \bibnamefont
  {Chernodub}}, \bibinfo {author} {\bibfnamefont {K.}~\bibnamefont {Ishiguro}},
  \bibinfo {author} {\bibfnamefont {Y.}~\bibnamefont {Mori}}, \bibinfo {author}
  {\bibfnamefont {Y.}~\bibnamefont {Nakamura}}, \bibinfo {author}
  {\bibfnamefont {M.~I.}\ \bibnamefont {Polikarpov}}, \bibinfo {author}
  {\bibfnamefont {T.}~\bibnamefont {Sekido}}, \bibinfo {author} {\bibfnamefont
  {T.}~\bibnamefont {Suzuki}}, \ and\ \bibinfo {author} {\bibfnamefont {V.~I.}\
  \bibnamefont {Zakharov}},\ }\href {\doibase 10.1103/PhysRevD.72.074505}
  {\bibfield  {journal} {\bibinfo  {journal} {Phys. Rev. D}\ }\textbf {\bibinfo
  {volume} {72}},\ \bibinfo {pages} {074505} (\bibinfo {year} {2005})},\
  \Eprint {http://arxiv.org/abs/hep-lat/0508004} {arXiv:hep-lat/0508004}
  \BibitemShut {NoStop}%
\bibitem [{\citenamefont {Suzuki}\ \emph {et~al.}(2009)\citenamefont {Suzuki},
  \citenamefont {Hasegawa}, \citenamefont {Ishiguro}, \citenamefont {Koma},\
  and\ \citenamefont {Sekido}}]{Suzuki:2009xy}%
  \BibitemOpen
  \bibfield  {author} {\bibinfo {author} {\bibfnamefont {T.}~\bibnamefont
  {Suzuki}}, \bibinfo {author} {\bibfnamefont {M.}~\bibnamefont {Hasegawa}},
  \bibinfo {author} {\bibfnamefont {K.}~\bibnamefont {Ishiguro}}, \bibinfo
  {author} {\bibfnamefont {Y.}~\bibnamefont {Koma}}, \ and\ \bibinfo {author}
  {\bibfnamefont {T.}~\bibnamefont {Sekido}},\ }\href {\doibase
  10.1103/PhysRevD.80.054504} {\bibfield  {journal} {\bibinfo  {journal} {Phys.
  Rev. D}\ }\textbf {\bibinfo {volume} {80}},\ \bibinfo {pages} {054504}
  (\bibinfo {year} {2009})},\ \Eprint {http://arxiv.org/abs/0907.0583}
  {arXiv:0907.0583 [hep-lat]} \BibitemShut {NoStop}%
\bibitem [{\citenamefont {Cea}\ \emph {et~al.}(2012)\citenamefont {Cea},
  \citenamefont {Cosmai},\ and\ \citenamefont {Papa}}]{Cea:2012qw}%
  \BibitemOpen
  \bibfield  {author} {\bibinfo {author} {\bibfnamefont {P.}~\bibnamefont
  {Cea}}, \bibinfo {author} {\bibfnamefont {L.}~\bibnamefont {Cosmai}}, \ and\
  \bibinfo {author} {\bibfnamefont {A.}~\bibnamefont {Papa}},\ }\href {\doibase
  10.1103/PhysRevD.86.054501} {\bibfield  {journal} {\bibinfo  {journal} {Phys.
  Rev. D}\ }\textbf {\bibinfo {volume} {86}},\ \bibinfo {pages} {054501}
  (\bibinfo {year} {2012})},\ \Eprint {http://arxiv.org/abs/1208.1362}
  {arXiv:1208.1362 [hep-lat]} \BibitemShut {NoStop}%
\bibitem [{\citenamefont {Abrikosov}(1957)}]{Abrikosov:1956sx}%
  \BibitemOpen
  \bibfield  {author} {\bibinfo {author} {\bibfnamefont {A.~A.}\ \bibnamefont
  {Abrikosov}},\ }\href@noop {} {\bibfield  {journal} {\bibinfo  {journal}
  {Sov. Phys. JETP}\ }\textbf {\bibinfo {volume} {5}},\ \bibinfo {pages} {1174}
  (\bibinfo {year} {1957})}\BibitemShut {NoStop}%
\bibitem [{\citenamefont {Nielsen}\ and\ \citenamefont
  {Olesen}(1973)}]{Nielsen:1973cs}%
  \BibitemOpen
  \bibfield  {author} {\bibinfo {author} {\bibfnamefont {H.~B.}\ \bibnamefont
  {Nielsen}}\ and\ \bibinfo {author} {\bibfnamefont {P.}~\bibnamefont
  {Olesen}},\ }\href {\doibase 10.1016/0550-3213(73)90350-7} {\bibfield
  {journal} {\bibinfo  {journal} {Nucl. Phys. B}\ }\textbf {\bibinfo {volume}
  {61}},\ \bibinfo {pages} {45} (\bibinfo {year} {1973})}\BibitemShut {NoStop}%
\bibitem [{\citenamefont {Witten}(1985)}]{Witten:1984eb}%
  \BibitemOpen
  \bibfield  {author} {\bibinfo {author} {\bibfnamefont {E.}~\bibnamefont
  {Witten}},\ }\href {\doibase 10.1016/0550-3213(85)90022-7} {\bibfield
  {journal} {\bibinfo  {journal} {Nucl. Phys. B}\ }\textbf {\bibinfo {volume}
  {249}},\ \bibinfo {pages} {557} (\bibinfo {year} {1985})}\BibitemShut
  {NoStop}%
\bibitem [{\citenamefont {Kaneko}\ \emph {et~al.}(2014)\citenamefont {Kaneko},
  \citenamefont {Aoki}, \citenamefont {Cossu}, \citenamefont {Fukaya},
  \citenamefont {Hashimoto},\ and\ \citenamefont {Noaki}}]{Kaneko:2013jla}%
  \BibitemOpen
  \bibfield  {author} {\bibinfo {author} {\bibfnamefont {T.}~\bibnamefont
  {Kaneko}}, \bibinfo {author} {\bibfnamefont {S.}~\bibnamefont {Aoki}},
  \bibinfo {author} {\bibfnamefont {G.}~\bibnamefont {Cossu}}, \bibinfo
  {author} {\bibfnamefont {H.}~\bibnamefont {Fukaya}}, \bibinfo {author}
  {\bibfnamefont {S.}~\bibnamefont {Hashimoto}}, \ and\ \bibinfo {author}
  {\bibfnamefont {J.}~\bibnamefont {Noaki}} (\bibinfo {collaboration}
  {JLQCD}),\ }\href {\doibase 10.22323/1.187.0125} {\bibfield  {journal}
  {\bibinfo  {journal} {PoS}\ }\textbf {\bibinfo {volume} {LATTICE2013}},\
  \bibinfo {pages} {125} (\bibinfo {year} {2014})},\ \Eprint
  {http://arxiv.org/abs/1311.6941} {arXiv:1311.6941 [hep-lat]} \BibitemShut
  {NoStop}%
\bibitem [{\citenamefont {Iritani}\ \emph {et~al.}(2013)\citenamefont
  {Iritani}, \citenamefont {Cossu},\ and\ \citenamefont
  {Hashimoto}}]{Iritani:2013apa}%
  \BibitemOpen
  \bibfield  {author} {\bibinfo {author} {\bibfnamefont {T.}~\bibnamefont
  {Iritani}}, \bibinfo {author} {\bibfnamefont {G.}~\bibnamefont {Cossu}}, \
  and\ \bibinfo {author} {\bibfnamefont {S.}~\bibnamefont {Hashimoto}},\ }\href
  {\doibase 10.22323/1.205.0159} {\bibfield  {journal} {\bibinfo  {journal}
  {PoS}\ }\textbf {\bibinfo {volume} {Hadron2013}},\ \bibinfo {pages} {159}
  (\bibinfo {year} {2013})},\ \Eprint {http://arxiv.org/abs/1401.4293}
  {arXiv:1401.4293 [hep-lat]} \BibitemShut {NoStop}%
\bibitem [{\citenamefont {Iritani}\ \emph
  {et~al.}(2014{\natexlab{a}})\citenamefont {Iritani}, \citenamefont {Cossu},\
  and\ \citenamefont {Hashimoto}}]{Iritani:2013rla}%
  \BibitemOpen
  \bibfield  {author} {\bibinfo {author} {\bibfnamefont {T.}~\bibnamefont
  {Iritani}}, \bibinfo {author} {\bibfnamefont {G.}~\bibnamefont {Cossu}}, \
  and\ \bibinfo {author} {\bibfnamefont {S.}~\bibnamefont {Hashimoto}},\ }\href
  {\doibase 10.22323/1.187.0376} {\bibfield  {journal} {\bibinfo  {journal}
  {PoS}\ }\textbf {\bibinfo {volume} {LATTICE2013}},\ \bibinfo {pages} {376}
  (\bibinfo {year} {2014}{\natexlab{a}})},\ \Eprint
  {http://arxiv.org/abs/1311.0218} {arXiv:1311.0218 [hep-lat]} \BibitemShut
  {NoStop}%
\bibitem [{\citenamefont {Iritani}\ \emph
  {et~al.}(2014{\natexlab{b}})\citenamefont {Iritani}, \citenamefont {Cossu},\
  and\ \citenamefont {Hashimoto}}]{Iritani:2014fga}%
  \BibitemOpen
  \bibfield  {author} {\bibinfo {author} {\bibfnamefont {T.}~\bibnamefont
  {Iritani}}, \bibinfo {author} {\bibfnamefont {G.}~\bibnamefont {Cossu}}, \
  and\ \bibinfo {author} {\bibfnamefont {S.}~\bibnamefont {Hashimoto}},\ }\href
  {\doibase 10.22323/1.214.0338} {\bibfield  {journal} {\bibinfo  {journal}
  {PoS}\ }\textbf {\bibinfo {volume} {LATTICE2014}},\ \bibinfo {pages} {338}
  (\bibinfo {year} {2014}{\natexlab{b}})},\ \Eprint
  {http://arxiv.org/abs/1412.2322} {arXiv:1412.2322 [hep-lat]} \BibitemShut
  {NoStop}%
\bibitem [{\citenamefont {Iritani}\ \emph {et~al.}(2015)\citenamefont
  {Iritani}, \citenamefont {Cossu},\ and\ \citenamefont
  {Hashimoto}}]{Iritani:2015zwa}%
  \BibitemOpen
  \bibfield  {author} {\bibinfo {author} {\bibfnamefont {T.}~\bibnamefont
  {Iritani}}, \bibinfo {author} {\bibfnamefont {G.}~\bibnamefont {Cossu}}, \
  and\ \bibinfo {author} {\bibfnamefont {S.}~\bibnamefont {Hashimoto}},\ }\href
  {\doibase 10.1103/PhysRevD.91.094501} {\bibfield  {journal} {\bibinfo
  {journal} {Phys. Rev. D}\ }\textbf {\bibinfo {volume} {91}},\ \bibinfo
  {pages} {094501} (\bibinfo {year} {2015})},\ \Eprint
  {http://arxiv.org/abs/1502.04845} {arXiv:1502.04845 [hep-lat]} \BibitemShut
  {NoStop}%
\bibitem [{\citenamefont {Iritani}\ \emph {et~al.}(2016)\citenamefont
  {Iritani}, \citenamefont {Cossu},\ and\ \citenamefont
  {Hashimoto}}]{Iritani:2016fvi}%
  \BibitemOpen
  \bibfield  {author} {\bibinfo {author} {\bibfnamefont {T.}~\bibnamefont
  {Iritani}}, \bibinfo {author} {\bibfnamefont {G.}~\bibnamefont {Cossu}}, \
  and\ \bibinfo {author} {\bibfnamefont {S.}~\bibnamefont {Hashimoto}},\ }\href
  {\doibase 10.1063/1.4938718} {\bibfield  {journal} {\bibinfo  {journal} {AIP
  Conf. Proc.}\ }\textbf {\bibinfo {volume} {1701}},\ \bibinfo {pages} {100009}
  (\bibinfo {year} {2016})}\BibitemShut {NoStop}%
\bibitem [{\citenamefont {Coleman}(1975)}]{Coleman:1974bu}%
  \BibitemOpen
  \bibfield  {author} {\bibinfo {author} {\bibfnamefont {S.~R.}\ \bibnamefont
  {Coleman}},\ }\href {\doibase 10.1103/PhysRevD.11.2088} {\bibfield  {journal}
  {\bibinfo  {journal} {Phys. Rev. D}\ }\textbf {\bibinfo {volume} {11}},\
  \bibinfo {pages} {2088} (\bibinfo {year} {1975})}\BibitemShut {NoStop}%
\bibitem [{\citenamefont {Mandelstam}(1975)}]{Mandelstam:1975hb}%
  \BibitemOpen
  \bibfield  {author} {\bibinfo {author} {\bibfnamefont {S.}~\bibnamefont
  {Mandelstam}},\ }\href {\doibase 10.1103/PhysRevD.11.3026} {\bibfield
  {journal} {\bibinfo  {journal} {Phys. Rev. D}\ }\textbf {\bibinfo {volume}
  {11}},\ \bibinfo {pages} {3026} (\bibinfo {year} {1975})}\BibitemShut
  {NoStop}%
\bibitem [{\citenamefont {Cea}\ and\ \citenamefont
  {Cosmai}(1995)}]{Cea:1995zt}%
  \BibitemOpen
  \bibfield  {author} {\bibinfo {author} {\bibfnamefont {P.}~\bibnamefont
  {Cea}}\ and\ \bibinfo {author} {\bibfnamefont {L.}~\bibnamefont {Cosmai}},\
  }\href {\doibase 10.1103/PhysRevD.52.5152} {\bibfield  {journal} {\bibinfo
  {journal} {Phys. Rev. D}\ }\textbf {\bibinfo {volume} {52}},\ \bibinfo
  {pages} {5152} (\bibinfo {year} {1995})},\ \Eprint
  {http://arxiv.org/abs/hep-lat/9504008} {arXiv:hep-lat/9504008} \BibitemShut
  {NoStop}%
\bibitem [{\citenamefont {Cardaci}\ \emph {et~al.}(2011)\citenamefont
  {Cardaci}, \citenamefont {Cea}, \citenamefont {Cosmai}, \citenamefont
  {Falcone},\ and\ \citenamefont {Papa}}]{Cardaci:2010tb}%
  \BibitemOpen
  \bibfield  {author} {\bibinfo {author} {\bibfnamefont {M.~S.}\ \bibnamefont
  {Cardaci}}, \bibinfo {author} {\bibfnamefont {P.}~\bibnamefont {Cea}},
  \bibinfo {author} {\bibfnamefont {L.}~\bibnamefont {Cosmai}}, \bibinfo
  {author} {\bibfnamefont {R.}~\bibnamefont {Falcone}}, \ and\ \bibinfo
  {author} {\bibfnamefont {A.}~\bibnamefont {Papa}},\ }\href {\doibase
  10.1103/PhysRevD.83.014502} {\bibfield  {journal} {\bibinfo  {journal} {Phys.
  Rev. D}\ }\textbf {\bibinfo {volume} {83}},\ \bibinfo {pages} {014502}
  (\bibinfo {year} {2011})},\ \Eprint {http://arxiv.org/abs/1011.5803}
  {arXiv:1011.5803 [hep-lat]} \BibitemShut {NoStop}%
\bibitem [{\citenamefont {Schmidt}(1907)}]{Schmidt:1907pp}%
  \BibitemOpen
  \bibfield  {author} {\bibinfo {author} {\bibfnamefont {E.}~\bibnamefont
  {Schmidt}},\ }\href {\doibase 10.1007/BF01449770} {\bibfield  {journal}
  {\bibinfo  {journal} {Mathematische Annalen}\ }\textbf {\bibinfo {volume}
  {63}},\ \bibinfo {pages} {433} (\bibinfo {year} {1907})}\BibitemShut
  {NoStop}%
\bibitem [{\citenamefont {Peres}(1995)}]{peres1995quantum}%
  \BibitemOpen
  \bibfield  {author} {\bibinfo {author} {\bibfnamefont {A.}~\bibnamefont
  {Peres}},\ }\href {https://books.google.co.jp/books?id=rMGqMyFBcL8C} {\emph
  {\bibinfo {title} {Quantum Theory: Concepts and Methods}}},\ Fundamental
  Theories of Physics\ (\bibinfo  {publisher} {Springer Netherlands},\ \bibinfo
  {year} {1995})\BibitemShut {NoStop}%
\bibitem [{\citenamefont {Baker}\ \emph {et~al.}(2020)\citenamefont {Baker},
  \citenamefont {Cea}, \citenamefont {Chelnokov}, \citenamefont {Cosmai},
  \citenamefont {Cuteri},\ and\ \citenamefont {Papa}}]{Baker:2019gsi}%
  \BibitemOpen
  \bibfield  {author} {\bibinfo {author} {\bibfnamefont {M.}~\bibnamefont
  {Baker}}, \bibinfo {author} {\bibfnamefont {P.}~\bibnamefont {Cea}}, \bibinfo
  {author} {\bibfnamefont {V.}~\bibnamefont {Chelnokov}}, \bibinfo {author}
  {\bibfnamefont {L.}~\bibnamefont {Cosmai}}, \bibinfo {author} {\bibfnamefont
  {F.}~\bibnamefont {Cuteri}}, \ and\ \bibinfo {author} {\bibfnamefont
  {A.}~\bibnamefont {Papa}},\ }\href {\doibase 10.1140/epjc/s10052-020-8077-5}
  {\bibfield  {journal} {\bibinfo  {journal} {Eur. Phys. J. C}\ }\textbf
  {\bibinfo {volume} {80}},\ \bibinfo {pages} {514} (\bibinfo {year} {2020})},\
  \Eprint {http://arxiv.org/abs/1912.04739} {arXiv:1912.04739 [hep-lat]}
  \BibitemShut {NoStop}%
\bibitem [{\citenamefont {Clem}(1975)}]{Clem:1975jr}%
  \BibitemOpen
  \bibfield  {author} {\bibinfo {author} {\bibfnamefont {J.~R.}\ \bibnamefont
  {Clem}},\ }\href {\doibase 10.1007/BF00116134} {\bibfield  {journal}
  {\bibinfo  {journal} {Journal of Low Temperature Physics}\ }\textbf {\bibinfo
  {volume} {18}},\ \bibinfo {pages} {427} (\bibinfo {year} {1975})}\BibitemShut
  {NoStop}%
\bibitem [{\citenamefont {Hasenfratz}\ \emph {et~al.}(1981)\citenamefont
  {Hasenfratz}, \citenamefont {Hasenfratz},\ and\ \citenamefont
  {Hasenfratz}}]{Hasenfratz:1980ue}%
  \BibitemOpen
  \bibfield  {author} {\bibinfo {author} {\bibfnamefont {A.}~\bibnamefont
  {Hasenfratz}}, \bibinfo {author} {\bibfnamefont {E.}~\bibnamefont
  {Hasenfratz}}, \ and\ \bibinfo {author} {\bibfnamefont {P.}~\bibnamefont
  {Hasenfratz}},\ }\href {\doibase 10.1016/0550-3213(81)90426-0} {\bibfield
  {journal} {\bibinfo  {journal} {Nucl. Phys. B}\ }\textbf {\bibinfo {volume}
  {180}},\ \bibinfo {pages} {353} (\bibinfo {year} {1981})}\BibitemShut
  {NoStop}%
\bibitem [{\citenamefont {Luscher}\ \emph {et~al.}(1981)\citenamefont
  {Luscher}, \citenamefont {Munster},\ and\ \citenamefont
  {Weisz}}]{Luscher:1980iy}%
  \BibitemOpen
  \bibfield  {author} {\bibinfo {author} {\bibfnamefont {M.}~\bibnamefont
  {Luscher}}, \bibinfo {author} {\bibfnamefont {G.}~\bibnamefont {Munster}}, \
  and\ \bibinfo {author} {\bibfnamefont {P.}~\bibnamefont {Weisz}},\ }\href
  {\doibase 10.1016/0550-3213(81)90151-6} {\bibfield  {journal} {\bibinfo
  {journal} {Nucl. Phys. B}\ }\textbf {\bibinfo {volume} {180}},\ \bibinfo
  {pages} {1} (\bibinfo {year} {1981})}\BibitemShut {NoStop}%
\bibitem [{\citenamefont {Cardoso}\ \emph {et~al.}(2013)\citenamefont
  {Cardoso}, \citenamefont {Cardoso},\ and\ \citenamefont
  {Bicudo}}]{Cardoso:2013lla}%
  \BibitemOpen
  \bibfield  {author} {\bibinfo {author} {\bibfnamefont {N.}~\bibnamefont
  {Cardoso}}, \bibinfo {author} {\bibfnamefont {M.}~\bibnamefont {Cardoso}}, \
  and\ \bibinfo {author} {\bibfnamefont {P.}~\bibnamefont {Bicudo}},\ }\href
  {\doibase 10.1103/PhysRevD.88.054504} {\bibfield  {journal} {\bibinfo
  {journal} {Phys. Rev. D}\ }\textbf {\bibinfo {volume} {88}},\ \bibinfo
  {pages} {054504} (\bibinfo {year} {2013})},\ \Eprint
  {http://arxiv.org/abs/1302.3633} {arXiv:1302.3633 [hep-lat]} \BibitemShut
  {NoStop}%
\bibitem [{\citenamefont {Bakry}\ \emph {et~al.}(2012)\citenamefont {Bakry},
  \citenamefont {Leinweber},\ and\ \citenamefont {Williams}}]{Bakry:2010sp}%
  \BibitemOpen
  \bibfield  {author} {\bibinfo {author} {\bibfnamefont {A.~S.}\ \bibnamefont
  {Bakry}}, \bibinfo {author} {\bibfnamefont {D.~B.}\ \bibnamefont
  {Leinweber}}, \ and\ \bibinfo {author} {\bibfnamefont {A.~G.}\ \bibnamefont
  {Williams}},\ }\href {\doibase 10.1103/PhysRevD.85.034504} {\bibfield
  {journal} {\bibinfo  {journal} {Phys. Rev. D}\ }\textbf {\bibinfo {volume}
  {85}},\ \bibinfo {pages} {034504} (\bibinfo {year} {2012})},\ \Eprint
  {http://arxiv.org/abs/1011.1380} {arXiv:1011.1380 [hep-lat]} \BibitemShut
  {NoStop}%
\end{thebibliography}%
\newpage

\end{document}